\documentclass[aps,pra,amssymb,amsfonts,twocolumn,superscriptaddress,showpacs]{revtex4}
\usepackage[stable]{footmisc}
\usepackage{array}
\usepackage{verbatim}
\usepackage{tikz}
\usepackage{hyperref}
\usepackage{color}

\usepackage{graphicx}
\usepackage{dcolumn}
\usepackage{bm}

\begin{document}

\title{A high-power 626 nm diode laser system for Beryllium ion trapping}%
\author{H. Ball}%
\author{M. W. Lee}%
\author{S. D. Gensemer}%
\altaffiliation{Present Address: Commonwealth Scientific and Industrial Research Organisation, Materials Science and Engineering, West Lindfield, NSW 2070, Australia}

\author{M.J. Biercuk}%
\email{michael.biercuk@sydney.edu.au}
\affiliation{ARC Centre for Engineered Quantum Systems, School of Physics, The
University of Sydney, NSW 2006 Australia}
\affiliation{National Measurement Institute, West Lindfield, NSW 2070 Australia}
\date{today}%
\begin{abstract}
We describe a high-power, frequency-tunable, external cavity diode laser (ECDL) system near 626nm useful for laser cooling of trapped $^9$Be$^+$ ions. A commercial single-mode laser diode with rated power output of 170 mW at 635nm is cooled to $\approx - 31$ C, and a single longitudinal mode is selected via the Littrow configuration.  In our setup, involving multiple stages of thermoelectric cooling, we are able to obtain $\approx$130 mW near 626 nm, sufficient for efficient frequency doubling to the required Doppler cooling wavelengths near 313nm in ionized Beryllium.   In order to improve nonlinear frequency conversion efficiency, we achieve larger useful power via injection locking of a slave laser.  In this way the entirety of the slave output power is available for frequency doubling, while analysis may be performed on the master output.  We believe that this simple laser system addresses a key need in the ion trapping community and dramatically reduces the cost and complexity associated with Beryllium ion trapping experiments.
\end{abstract}

\maketitle

\section{Introduction}

Over the past several decades trapped ions have emerged as a powerful architecture for breakthrough research into quantum information, computation and simulation\cite{James1998,Blatt2008,Wineland2009}. Key developments include proposals and experiments to implement quantum logic gates~\cite{Zoller1995, Wineland1995}, exceptionally long coherence times~\cite{Blatt2008}, and high-fidelity operations~\cite{Wineland1998,Wineland2002,Wineland2010,Wineland2011}as needed in large-scale quantum computation. Among the wide variety of trapped ion species in use, Beryllium provides significant advantages~\cite{Wineland1998} to the experimentalist: the light mass of $^9$Be$^+$ results in high motional frequencies in RF pseudopotential traps, the low atomic number provides a relatively simple level structure,  the presence of strong \emph{cycling transition} enables high-fidelity state detection and efficient cooling, and the presence of field-insensitive \emph{clock} transitions can provide long-lived qubit coherences (10s of seconds)~\cite{Langer2006}. A large number of interesting experiments leveraging these capabilities has been conducted using $^9$Be$^+$, ranging from quantum computing and quantum simulation to precision frequency metrology and sensing~\cite{Bollinger1991,Kielpinski2003,Wineland1998,Monroe2002, King1999,DidiGate, Langer2006,Jost2010,Biercuk2009I,Biercuk2012, Sawyer2012}. 

Despite these advantages, the technical complexity of laser systems required for $^9$Be$^+$ trapping has reduced the uptake of this species in new experimental laboratories.  The primary transitions in Beryllium near 313 nm mandate tunable, high-power, narrow-linewidth, low-drift CW lasers in a wavelength range that is not well covered by commercial products.  For instance, semiconductor diode lasers - appreciated for their relatively low technical complexity and cost - do not extend appreciably below $\sim$370 nm using GaN-based diodes.  A traditional approach to generating the colors required for experiments with $^9$Be$^+$ has been to frequency double ~100-500 mW red output from a ring dye laser operating near 626 nm.  More recent work has seen the development of all-solid-state sources using infrared fiber lasers and nonlinear frequency conversion~\cite{Schiller2002,Hansch2006,Schiller2011,Wilson2011}. Both approaches, however, are costly and technically complex and a compact, economical alternative leveraging semiconductor diode lasers is highly desirable. 

In this manuscript we describe the design and construction of a compact, economic external cavity diode laser (ECDL) system outputting 140 mW red light near 626 nm - sufficient for subsequent frequency doubling to produce $\sim3-5$ mW ($\sim$70 mW red input to cavity after system losses) near 313nm using a home-built doubling cavity~\cite{Wilson2011}.   This work leverages recent advances in the production of high-power single-mode semiconductor laser diodes operating near 635nm which may be cryogenically cooled to shift their gain profiles near 626 nm at $T\approx -31^{\circ}$C.   The laser is assembled in a hermetic enclosure in the Littrow configuration in order to establish a frequency selective external cavity.  Mode-hop free piezo tuning of $\sim$5 GHz is achieved allowing broad tuning across spectral features of interest for Beryllium experiments.  Long-term stability is provided by stabilizing the laser output to a hyperfine transition in molecular iodine or direct feedback from a commercial wavemeter.  We also perform injection locking using two diodes in master-slave configuration in order to produce higher usable red-light output; the slave's full power may be directed to frequency doubling with analysis and frequency stabilization performed directly on the master laser.  This development represents a major simplification of the requisite laser systems for the $^9$Be$^+$ ion-trapping community and should expand the accessibility of this species to a broader range of researchers.

The remainder of this manuscript is organized as follows.  In Section~\ref{Sec:ECDL} we outline the basic operating principles of the external cavity diode laser and temperature tuning.  This is followed by a detailed description of our experimental system in Section~\ref{Sec:Expt}.  We then describe the system's performance relating to laser frequency output and injection locking in Section~\ref{Sec:Operation}, followed by a conclusion. 

\section{Temperature-tuned external cavity diode lasers\label{Sec:ECDL}}

In a semiconductor diode laser a double $p-n$ hetereojunction acts both as the gain medium and as a waveguide to form an optical resonator for lasing. An injection current produces a population inversion by `injecting' electrons and holes from opposite sides of the junction into the depletion region, enabling them to radiatively recombine across the energy bandgap $E_g$ via stimulated emisison of photons, which are emitted through the edges of the semiconductor layers\cite{Saleh1991, Siegman1986}.  Free-running index-guided laser diodes have multiple longitudinal cavity modes, typically separated by 100-200 GHz, within  a gain profile with width~$\sim$10 nm\cite{Ricci1995}.   Determination of the ultimate lasing wavelength thus requires consideration of the interplay between these two physical effects.   A result of this competition is a large free-running linewidth, typically of order nm.

The peak wavelength of the diode's gain profile is primarily determined by the bandgap, $E_{g}$, of the semiconductor junction.  A user-accessible parameter influencing this value \emph{in situ} is the temperature of the diode.  In particular, the temperature dependence of the bandgap derives from both lattice expansion and electron-phonon interaction~\cite{Fan1951}.  Both contributions are proportional to the phonon occupancy of the lattice mode at frequency $\omega$, $n(\omega,T)\propto\coth(\hbar\omega/2kT)$, with $k$ the Boltzmann constant. A number of semi-empirical formulae (c.f. \cite{Varshni1967}) have been devised to model this behaviour, including a relatively simple three-parameter model proposed by Chen \emph{et al.}\cite{Chen1991} 
\begin{eqnarray}
E_{g}(T)=E_{g}(0)-S\langle\hbar\omega\rangle[\coth(\langle\hbar\omega\rangle/2kT)-1].
\label{Bandgap}
\end{eqnarray}
Here $E_{g}(0)$ is the bandgap at zero temperature, $S$ is a dimensionless coupling constant, and $\langle\hbar\omega\rangle$ is the average phonon energy with angle brackets indicating an ensemble average over all phonon modes.  The primary observation derived from this formalism is that the maximum gain of the medium can be temperature tuned;  decreasing the temperature shifts the gain profile to lower wavelengths, typically with a change of $\sim$0.25 nm K$^{-1}$~\cite{Wieman1991}.  

Once the gain profile is established in a particular wavelength range, the competition between longitudinal modes within the gain profile determines the emission wavelength, with the laser operating at the modes which experience the maximum gain.  These may be shifted by tuning the optical cavity length via the injection current, which changes both the instantaneous temperature and the carrier density in the junction.  However the spectral content of a laser diode can be dramatically improved, including reduction to single-mode operation, by exploiting their susceptibility to optical feedback~\cite{MacAdam1992, Wieman1991}.

\begin{figure}
\includegraphics[width=1\columnwidth]{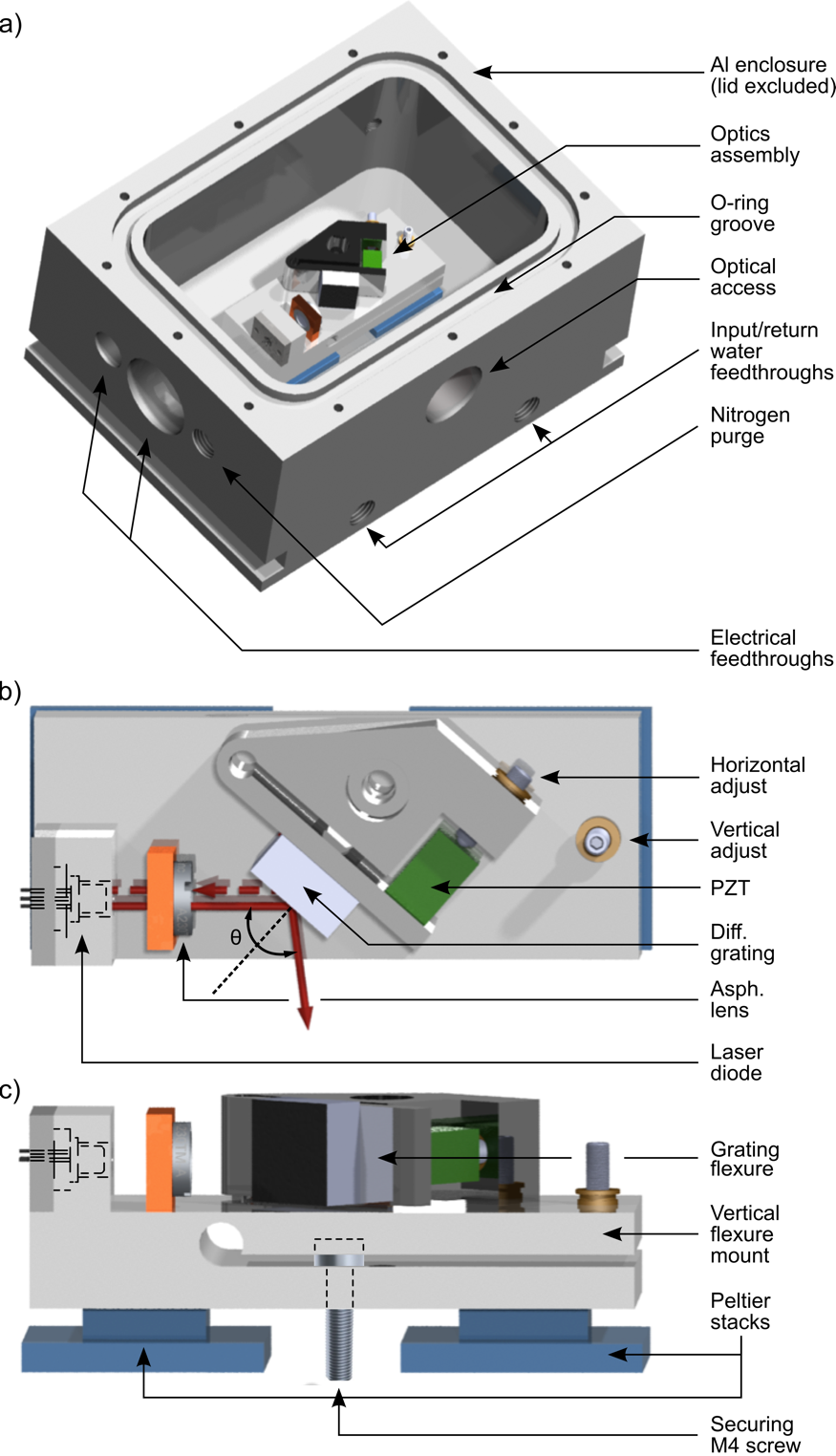}
\caption{Schematic of the laser diode assembly.  a) Complete assembly in hermetic enclosure.  b) and c) Top and side-views, respectively, of the diode mount assembly.  Diode is mounted in baseplate which is thermally coupled to Peltier cooling stacks. Slits in baseplate and grating mount provide vertical and horizontal flexure adjustment under tension from micrometer screws. Grating is mounted in Littrow configuration to reflect the first-order diffracted light directly back into diode chip, indicated by red-dashed line. A similar ECDL assembly has been successfully used to operate cryogenically cooled 780 nm diodes, pulled to 767 nm for laser cooling of neutral K.  It has also been used for high-temperature operation of 660 nm diodes near 670 nm for laser cooling of Li.~\cite{Tiecke_thesis} } \label{ECDLassembly}
\end{figure}

Selecting a single longitudinal mode requires the establishment of an external cavity in order to provide optical feedback at a preferred wavelength.  This may be accomplished using reflection from a diffraction grating in the Littrow configuration, forming an external cavity between the grating and the rear facet of the diode.  The grating is tilted with respect to the incident beam from the laser diode, with the angle of incidence chosen such that the first-order diffraction is sent directly back into the laser diode while the zeroth order serves as the output beam (Fig. \ref{ECDLassembly}b), thus acting as a wavelength-dependent mirror~\cite{Ricci1995}.  The \emph{Littrow angle} providing this optical feedback may be written
\begin{eqnarray}
\theta_i=\sin^{-1}\frac{\lambda}{2d} = \sin^{-1}\frac{N \lambda}{2}.
\label{LittrowAngle}
\end{eqnarray}
where $d=1/N$ is the line-spacing on the grating given a line-density $N$, and $\lambda$ is the optical wavelength.  Thus tuning the grating angle provides wavelength selectivity in the external cavity.  From a practical perspective one selects a grating with line-density such that the zeroth order is reflected nearly perpendicular to the incident beam, as indicated by the grating equation~\cite{Hecht1974} for $m$-th order diffraction $d(\sin\theta_m+\sin\theta_i)=m\lambda$, given angle-of-incidence $\theta_{i}$.

\section{Experimental System}\label{Sec:Expt}
Our system incorporates a 170 mW, single-mode, GaAs-based laser diode from Opnext (model HL63133DG) or Mitsubishi (ML520G55 ) in a standard $\phi$ 5.6 mm package. 
These diodes operate at 635-637 nm at room temperature and have a measured gain-curve shift of $\sim$0.20 nm/K, determined from a linear fit of experimental data of wavelength dependence on temperature.  Accordingly, in order to pull the peak wavelength of the gain profile the required 11 nm to to 626 nm, we must cool the laser diode to near $-31 ^\circ$ C. 

Accomplishing this temperature tuning requires careful optical system design providing a hermetic enclosure and a thermo-electrically cooled laser diode mount~\cite{note1}.  The ECDL is housed in an airtight enclosure machined from a block of Al (Fig. \ref{ECDLassembly}a).  A groove is milled into the top of the enclosure to seat an O-ring between a plexiglass lid (not shown). All electrical connections are made with airtight connectors at feedthroughs in the enclosure. 

The diode and associated optics are assembled on a ($75\times30\times12$ mm) Aluminium flexure mount based on the design of Hansch \emph{et al.}\cite{Ricci1995} (hereafter, ``diode assembly'').  The main diode mount is integrated into this assembly, with a 5.6 mm bored hole to house the diode.  The diode has a wide ($\sim$ 15 $^\circ$) beam divergence angle which must be corrected by a collimating lens before feedback from the grating is possible. For this purpose we use an $f= 4.5$ mm aspeheric Rochester lens pre-mounted in a 9 mm threaded ring, installed in a 12 mm square lens mount which is glued in place on the diode asembly.

As shown in Fig. \ref{ECDLassembly}c the diode assembly is mounted on two stages of Peltier thermoelectric coolers (TEC) from TE technology, each stage consisting of a series pair of TECs.  The lower pair (item TE-127-1.0-1.3) is operated at a constant current, drawing 68 W from a benchtop power supply.  The upper pair (item TE-31-1.0-2.0), which draws 11.2 W, is driven by a Thorlabs temperature controller (item TED200C).  Feedback for temperature stabilization is accomplished using the voltage drop across a 3 k$\Omega$ NTC thermistor (not shown) placed in thermal contact with the diode mount, a few mm away from the laser diode itself.  

The diode assembly is attached to the floor of the enclosure by an M4 stainless steel screw to ensure  good thermal contact between the surfaces of the TEC stacks and both the baseplate and flexure mount. All thermal junctions within the enclosure, including between the diode package and the inner surface of the bore in which it is seated, are filled with a thin layer of low-outgassing cryogenic vacuum grease.  To maximize cooling efficiency we install thermal insulation made from a double layer of high-density foam sheeting placed around the diode assembly. The laser enclosure is water-cooled and thermally isolated from the thermal mass of the optical table.

Optical feedback is provided using a reflective holographic grating from Thorlabs (item GH13-24V) with a line density $N=2400$ mm$^{-1}$, giving a Littrow angle of 48.7 $^\circ$ at 626 nm.  As shown in Fig. \ref{ECDLassembly}b, the feedback-grating is cemented with Torr seal to an adjustable Al flexure mount, in turn secured to the vertical flexure mount using an M4 screw. The grating angle is coarsely adjusted by manual rotation, and fine adjustment is achieved using both a fine-thread micrometer screw and a piezoelectric actuator (PZT).  A similar micrometer screw in the vertical flexure mount  enables adjustment of the vertical angular displacement of the grating. 

The diode is oriented such that its polarization is parallel to the lines of the grating, providing stable optical feedback with optimal wavelength sensitivity~\cite{Ricci1995}.  This corresponds to orienting the major axis of the elliptical beam in the horizontal plane with about 20\% of the incident power coupled into the first diffraction order.

\section{Laser Operation and Characterization}\label{Sec:Operation}
\subsection{Alignment and cooling}
The lens position and flexure adjustors require tuning to optimize the grating position for the most efficient optical feedback. Prior to being cemented in place, the lens position is adjusted via a telescopic extension arm attached to a 3D translation stage. First alignment is performed at room temperature with the diode wavelength near 636 nm and a corresponding Littrow angle near $50^\circ$. Vertical alignment is accomplished by iteratively minimizing the threshold current.  With these diodes the free-running threshold current $\sim$75 mA is pulled to $\sim$50 mA when the alignment is correct.  

Before cooling the grating angle is coarsely adjusted to 48.7 $^\circ$ and the enclosure purged with a slow flow of nitrogen via a gas inlet in the side of the enclosure. This removes residual water vapour and prevents condensation on the optics after enclosure is sealed and the ECDL cooled.  The TECs are then energized and the diode is cooled to near -31$^\circ$ C and feedback controlled at this temperature.  Cooling water for the laser enclosure is set to 12$^\circ$ C, above the dewpoint in the laboratory.  This temperature setting minimizes the thermal gradient across the Peltiers while preventing condensation on the exterior of the enclosure.  With thermal insulation internal to the enclosure and between the enclosure and optical table as described above, we have successfully cooled the laser diode below -40$^{\circ}$C.  Additionally, we observe that the threshold current falls with reduced temperature, and while we have not confirmed this, diode lifetime generally increases. 
  
After cooling the diode we can select the desired longitudinal mode within the shifted gain profile near 626 nm appropriate for the Beryllium transitions. Threaded holes in the enclosure body, normally plugged with M6 screws permit access to the micrometer screws at low temperature as required for optimization of the optical alignment.  While the plugs are removed, we continuously purge the enclosure with dry nitrogen.  Once the grating is tuned, the enclosure is sealed and no further adjustment is required so long as the cryogenic temperature is maintained.

\begin{figure}[ht]
\includegraphics[width=1\columnwidth]{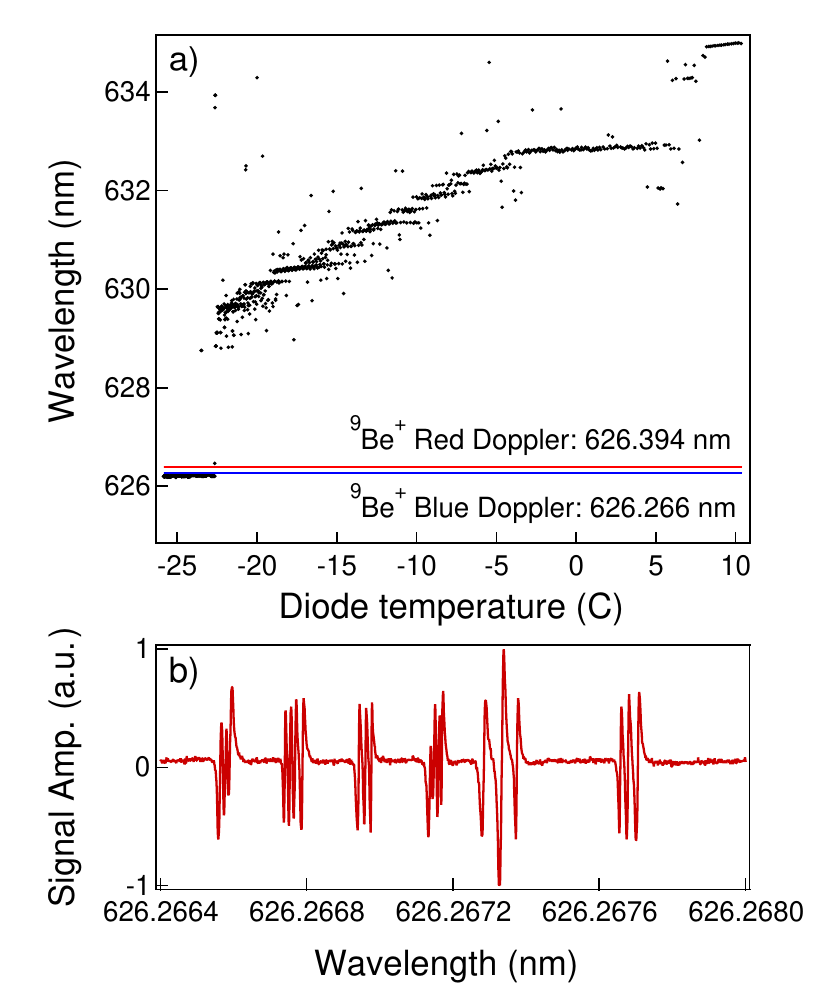}
\caption{Characterization of the output wavelength of the ECDL.  \textbf{a}. Temperature dependence of ECDL emission wavelength. LD is operating at 64 mW and grating is tuned to 626.266 nm. Black: emission wavelength measured by a high-finesse wavemeter.   Measurements taken as temperature falls (right-to-left).  Target wavelengths near 626 nm identified on the figure.   \textbf{b}. Sub-Doppler Iodine features near 626.266 nm obtained using frequency-modulated lock-in-detection and polarization spectroscopy on an Iodine vapour cell. Data were recorded from an oscilloscope monitoring the error signal generated using frequency-modulated lock-in-detection of the the iodine features while scanning the piezo at $\sim$10 Hz.   Wavelength was calibrated using direct measurements of feature location via the wavemeter. Approximately 5-10 mW of optical power was used for Iodine spectroscopy. }
\label{ECDLbehaviour}
\end{figure}

\subsection{Output characterization}
An ultra-low-noise laser diode controller from MOGLabs is employed to drive the injection current for the diode.  Normal operation for the selected diodes is up to 185 mA to allow operation safely below the maximum rated optical power.  It is vital to carefully monitor the operating current as the output power efficiency improves at low temperatures, but damage resistance does not.  Characterization of the output on a wavemeter or Fabry-Perot optical spectrum analyzer shows single-mode operation over a wide range of operating currents. 

Fig. \ref{ECDLbehaviour}a shows the behavior of the output wavelength from the ECDL as a function of temperature. The grating angle has been tuned to select a longitudinal mode near 626.266 nm (blue horizontal line), such that when frequency doubled the resulting UV is near 313.132 nm, appropriate for Doppler cooling of $^9$Be$^+$ near zero magnetic field.  For this data the output of the ECDL was coupled into an optical fiber and measured using a High-Finess WSU-10 wavemeter with 10 MHz absolute frequency accuracy.  An automated routine simultaneously collected measurements of thermistor resistance in order to measure diode temperature.  A calibration chart allowed conversion of resistance to absolute units of temperature. 

The peak wavelength of the gain profile decreases as we cool from room temerature to near -27 $^\circ$ C. At around -22$^\circ$ C the gain profile has migrated far enough that it begins to overlap with the longitudinal external cavity mode near 626.266. At this point optical feedback on that mode locks emission to the target wavelength.  We observe the ability to tune the output of the diode laser using optical feedback alone over a range of $\pm4$nm.

Once thermalized the laser wavelength may be finely adjusted via the piezo actuator.  Fig.~\ref{ECDLbehaviour}b shows the well-known sub-Doppler transitions of molecular iodine in the region of 626.266 nm~\cite{IodineAtlas}. Our measurement involves saturated polarization spectroscopy to give a dispersive lineshape~\cite{Ratnapala2004}, and reveals features familiar to Beryllium ion trap experimentalists~\cite{King1999}.  For instance, doppler cooling lasers are typically frequency stabilized using an offset lock a few hundred MHz red of the left-most visible transition.  We have successfully stabilized our laser to Iodine features using piezo regulation with long-term deviations less than 2 MHz over 1000s of seconds.  

We have measured the instantaneous linewidth of the ECDL by performing a heterodyne interference experiment, comparing the ECDL against an independent source of 626 nm light derived from narrow-linewidth ($\sim$100 kHz) fiber lasers and a nonlinear sum-frequency generation technique~\cite{Wilson2011}.  The instantaneous linewidth of the heterodyne beat note near 10 MHz was measured using a spectrum analyzer and found to be of order a few hundred kHz when the ECDL was operating near its maximum output power ($\sim$180 mA drive current).  This is achieved without any fast feedback on the diode current by virtue of the extremely low noise in the MOGLabs current source.

Notably, beam quality appears to vary considerably between the candidate diodes we have tested.  While our results are by no means exhaustive, we have found better quality output modes with Opnext diodes, allowing $>55\%$ throughput when coupling into an optical fiber, as opposed to a maximum of $\sim30\%$ when using Mitsubishi diodes.  We have also observed variations in output polarization homogeneity between diodes of a particular type.  Prescreening of candidate diodes at room temperature reveals these differences.



\subsection{Injection locking}

\begin{figure}[b]
\includegraphics[width=8cm]{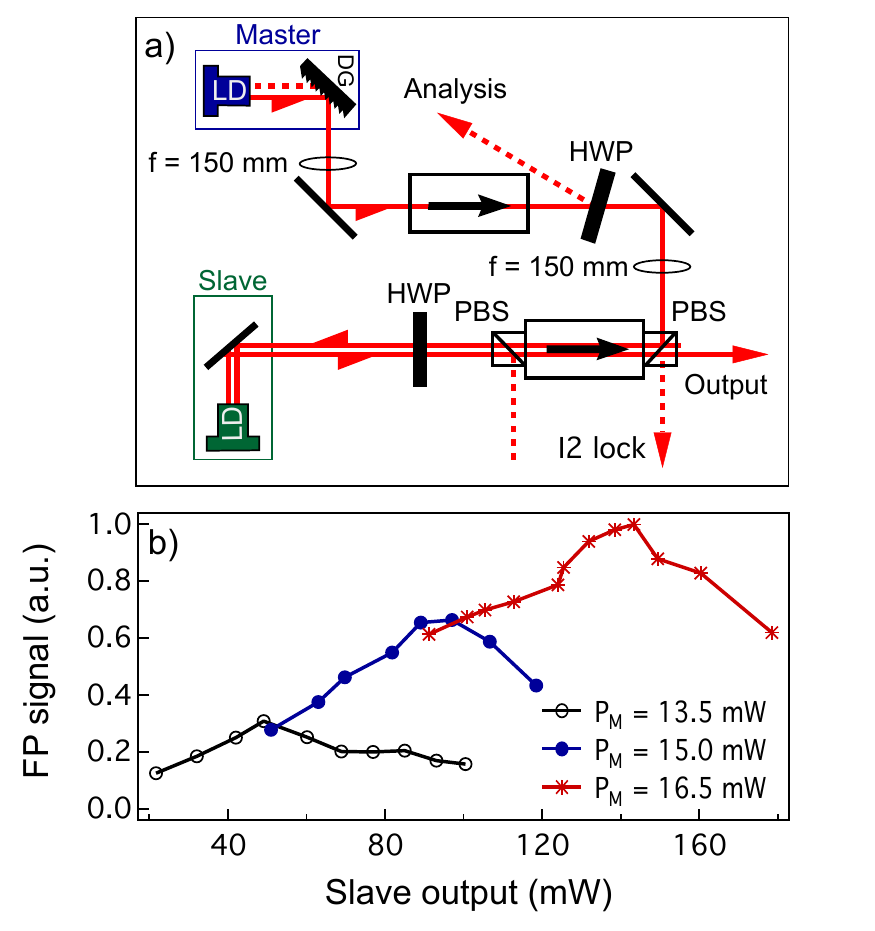}
\caption{Injection locking in a master-slave configuration.  \textbf{a} Schematic of the injection locking beamline.  The master light is focused with a 150 mm lens to pass through the aperture of an optical isolator (Thorlabs IO-3-633-LP) from which it emerges with a +45 $^\circ$ rotated polarization. This passes through a half wave plate (HWP) which restores the polarization to vertical, before being recollimated by a second lens and injected into the polarizing beamsplitter output of a second isolator (Thorlabs IO-5-633-PBS).  The reflected light provides the optical feedback to the slave laser.  A small reflected component from the first HWP is used for analysis.  \textbf{b} Characterization of optical output from the slave laser for different injected powers.  The size of a single optical mode measured using a Fabry-Perot spectrum analyzer grows linearly with slave output power until the injection power is insufficient to maintain single mode injection locking.}
\label{InjectionLock}
\end{figure}

In general Doppler cooling situations a few mW of UV is more than sufficient for practical purposes.  However, in cases where, e.g. Raman transitions are desired, higher optical powers are required~\cite{Wilson2011}, and the amount of available red light must be optimized.  In the direct ECDL configuration we extract a few tens of mW from the main beamline for wavemeter measurement, Iodine spectroscopy, Fabry-Perot stabilization, etc., reducing the achievable UV output.

An alternate arrangement can provide greater available red optical power by using an injection-locked master-slave configuration between two laser diodes.  In this situation we employ one laser configured as an ECDL and inject a small amount of light into a second diode.  The narrow-linewdith frequency stabilized output of the master provides optical feedback to the slave so long as the slave's gain profile is thermally tuned to the appropriate range.

We employ the beamline depicted schematically in Fig. \ref{InjectionLock}a. The master laser is an ECDL tuned to a longitudinal mode near 626 nm, with a vertically polarized output, and both master and slave diodes are oriented to output vertically polarized light. Alignment for injection locking is accomplished using an optical isolator with accessible polarizing beam cubes.   This isolator is oriented to transmit the slave's output at +45$^\circ$ linear polarization, following a half-wave plate.  The transmitted component of the master light at the PBS is used for Iodine spectroscopy and the reflected component is transmitted backward through the isolator.  This backward transmitted component emerges from the input side of the isolator at -45 $^\circ$ and then passes through the HWP which rotates its polarization back to vertical, matching it to the slave.  The master light is then reflected and focused in through the front facet of the slave diode via a collimating aspheric lens at its output.   A pair of 150mm lenses in the master beamline allows adjustment of the beam waist to maximize overlap of the spatial modes of the two diodes.   The injection-locked slave output power is transmitted through the optical isolator and emerges ready to be fully converted to UV.

Fig. \ref{InjectionLock} shows our characterization of the injection lock. The output from the slave (transmitted through the isolator) is sent to a Fabry-Perot (FP) intererometer for spectral characterization.  We seek single-mode operation in the optical spectrum analyzer and characterize the relative strength of the output mode for different injection powers.   The overall output power of the injection-locked slave is measured with an optical power meter.  The FP signal strength as a function of slave output power has been recorded for three different powers of the master laser: (13.5, 15.0, 16.5) mW. In each data set, single-mode operation is observed until the respective maxima are reached, at which points secondary optical modes emerge in the spectrograph.  In this regime the injected optical power from the master laser is insufficient to provide optical feedback to force the slave diode to operate single-mode. We observed successful injection locking up to $\sim 140$ mW output from the slave, the upper limit at which the diode should by safely operated.


\section{Conclusion}

We have detailed the design and operation of a frequency-tunable ECDL emitting red light near 626 nm.  Notably, the optical power of up to 140 mW, single-mode, is sufficient to enable efficient frequency conversion to the UV, providing access to the relevant cycling transitions in $^{9}$Be$^{+}$ ions.  Our approach leverages longstanding techniques in the community and recent advances in the availability of high-power single-mode laser diodes.  

The system we have developed produces up to $\sim$130 mW of reliable single-mode, narrow-linewidth optical power near 626 nm.  In an injection locked configuration we achieve up to 140 mW of light useful for frequency doubling to obtain approximately 5-7 mW of UV, and another $\sim$100 mW useful for analysis, frequency stabilization, etc.  This approach is extensible to multiple slave lasers permitting phase-coherent optical power near 626 nm suitable for high-power Raman laser experiments.  We believe that this technical development addresses a key issue in ion trapping experiments using Beryllium by reducing the cost and technical complexity associated with the required laser systems.

\begin{acknowledgments}
This work partially supported by the US Army Research Office under Contract Number  W911NF-11-1-0068, and the Australian Research Council Centre of Excellence for Engineered Quantum Systems CE110001013, and the Office of the Director of National Intelligence (ODNI), Intelligence Advanced Research Projects Activity (IARPA), through the Army Research Office. All statements of fact, opinion or conclusions contained herein are those of the authors and should not be construed as representing the official views or policies of IARPA, the ODNI, or the U.S. Government.
\end{acknowledgments}

\bibliography{ECDLresearch}

\begin{thebibliography}{36}
\expandafter\ifx\csname natexlab\endcsname\relax\def\natexlab#1{#1}\fi
\expandafter\ifx\csname bibnamefont\endcsname\relax
  \def\bibnamefont#1{#1}\fi
\expandafter\ifx\csname bibfnamefont\endcsname\relax
  \def\bibfnamefont#1{#1}\fi
\expandafter\ifx\csname citenamefont\endcsname\relax
  \def\citenamefont#1{#1}\fi
\expandafter\ifx\csname url\endcsname\relax
  \def\url#1{\texttt{#1}}\fi
\expandafter\ifx\csname urlprefix\endcsname\relax\def\urlprefix{URL }\fi
\providecommand{\bibinfo}[2]{#2}
\providecommand{\eprint}[2][]{\url{#2}}

\bibitem[{\citenamefont{James}(1998)}]{James1998}
\bibinfo{author}{\bibfnamefont{D.}~\bibnamefont{James}},
  \bibinfo{journal}{Applied Physics B: Lasers and Optics}
  \textbf{\bibinfo{volume}{66}}, \bibinfo{pages}{181} (\bibinfo{year}{1998}).

\bibitem[{\citenamefont{Wineland and Blatt}(2008)}]{Blatt2008}
\bibinfo{author}{\bibfnamefont{D.}~\bibnamefont{Wineland}} \bibnamefont{and}
  \bibinfo{author}{\bibfnamefont{R.}~\bibnamefont{Blatt}},
  \bibinfo{journal}{Nature} \textbf{\bibinfo{volume}{453}},
  \bibinfo{pages}{1008} (\bibinfo{year}{2008}).

\bibitem[{\citenamefont{Wineland and Leibfried}(2009)}]{Wineland2009}
\bibinfo{author}{\bibfnamefont{D.}~\bibnamefont{Wineland}} \bibnamefont{and}
  \bibinfo{author}{\bibfnamefont{D.}~\bibnamefont{Leibfried}},
  \bibinfo{journal}{Phys. Scr.} \textbf{\bibinfo{volume}{T137}},
  \bibinfo{pages}{014007} (\bibinfo{year}{2009}).

\bibitem[{\citenamefont{Cirac and Zoller}(1995)}]{Zoller1995}
\bibinfo{author}{\bibfnamefont{J.}~\bibnamefont{Cirac}} \bibnamefont{and}
  \bibinfo{author}{\bibfnamefont{P.}~\bibnamefont{Zoller}},
  \bibinfo{journal}{Phys. Rev. Lett.} \textbf{\bibinfo{volume}{74}},
  \bibinfo{pages}{4091} (\bibinfo{year}{1995}).

\bibitem[{\citenamefont{Monroe et~al.}(1995)\citenamefont{Monroe, Meekhof,
  King, Itano, and Wineland}}]{Wineland1995}
\bibinfo{author}{\bibfnamefont{C.}~\bibnamefont{Monroe}},
  \bibinfo{author}{\bibfnamefont{D.}~\bibnamefont{Meekhof}},
  \bibinfo{author}{\bibfnamefont{B.}~\bibnamefont{King}},
  \bibinfo{author}{\bibfnamefont{W.}~\bibnamefont{Itano}}, \bibnamefont{and}
  \bibinfo{author}{\bibfnamefont{D.}~\bibnamefont{Wineland}},
  \bibinfo{journal}{Phys. Rev. Lett.} \textbf{\bibinfo{volume}{75}},
  \bibinfo{pages}{4714} (\bibinfo{year}{1995}).

\bibitem[{\citenamefont{Wineland et~al.}(1998)\citenamefont{Wineland, Monroe,
  Itano, Leibfried, and King}}]{Wineland1998}
\bibinfo{author}{\bibfnamefont{D.}~\bibnamefont{Wineland}},
  \bibinfo{author}{\bibfnamefont{C.}~\bibnamefont{Monroe}},
  \bibinfo{author}{\bibfnamefont{W.}~\bibnamefont{Itano}},
  \bibinfo{author}{\bibfnamefont{D.}~\bibnamefont{Leibfried}},
  \bibnamefont{and} \bibinfo{author}{\bibfnamefont{B.~e.~a.}
  \bibnamefont{King}}, \bibinfo{journal}{J.Res.Natl.Inst.Stand.Tech.}
  \textbf{\bibinfo{volume}{103}}, \bibinfo{pages}{259} (\bibinfo{year}{1998}).

\bibitem[{\citenamefont{Wineland and
  Leibfried}(2011{\natexlab{a}})}]{Wineland2002}
\bibinfo{author}{\bibfnamefont{D.}~\bibnamefont{Wineland}} \bibnamefont{and}
  \bibinfo{author}{\bibfnamefont{D.}~\bibnamefont{Leibfried}},
  \bibinfo{journal}{Laser Phys. Lett.} \textbf{\bibinfo{volume}{8}},
  \bibinfo{pages}{175} (\bibinfo{year}{2011}{\natexlab{a}}).

\bibitem[{\citenamefont{Amini et~al.}(2010)\citenamefont{Amini, Uys, Wesenberg,
  Seidelin, Britton, Bollinger, Leibfried, Ospelkaus, A.P., and
  D.J.}}]{Wineland2010}
\bibinfo{author}{\bibfnamefont{J.}~\bibnamefont{Amini}},
  \bibinfo{author}{\bibfnamefont{H.}~\bibnamefont{Uys}},
  \bibinfo{author}{\bibfnamefont{J.}~\bibnamefont{Wesenberg}},
  \bibinfo{author}{\bibfnamefont{S.}~\bibnamefont{Seidelin}},
  \bibinfo{author}{\bibfnamefont{J.}~\bibnamefont{Britton}},
  \bibinfo{author}{\bibfnamefont{J.}~\bibnamefont{Bollinger}},
  \bibinfo{author}{\bibfnamefont{D.}~\bibnamefont{Leibfried}},
  \bibinfo{author}{\bibfnamefont{C.}~\bibnamefont{Ospelkaus}},
  \bibinfo{author}{\bibfnamefont{V.}~\bibnamefont{A.P.}}, \bibnamefont{and}
  \bibinfo{author}{\bibfnamefont{W.}~\bibnamefont{D.J.}}, \bibinfo{journal}{N.
  J. Phys.} \textbf{\bibinfo{volume}{12}}, \bibinfo{pages}{033031}
  (\bibinfo{year}{2010}).

\bibitem[{\citenamefont{Wineland and
  Leibfried}(2011{\natexlab{b}})}]{Wineland2011}
\bibinfo{author}{\bibfnamefont{D.}~\bibnamefont{Wineland}} \bibnamefont{and}
  \bibinfo{author}{\bibfnamefont{D.}~\bibnamefont{Leibfried}},
  \bibinfo{journal}{Laser Phys. Lett.} \textbf{\bibinfo{volume}{8}},
  \bibinfo{pages}{175} (\bibinfo{year}{2011}{\natexlab{b}}).

\bibitem[{\citenamefont{Langer}(2006)}]{Langer2006}
\bibinfo{author}{\bibfnamefont{C.}~\bibnamefont{Langer}}, \bibinfo{journal}{PhD
  thesis} \textbf{\bibinfo{volume}{University of Colorado}}
  (\bibinfo{year}{2006}).

\bibitem[{\citenamefont{Bollinger et~al.}(1991)\citenamefont{Bollinger, Heizen,
  Itano, Gilbert, and Wineland}}]{Bollinger1991}
\bibinfo{author}{\bibfnamefont{J.}~\bibnamefont{Bollinger}},
  \bibinfo{author}{\bibfnamefont{D.}~\bibnamefont{Heizen}},
  \bibinfo{author}{\bibfnamefont{W.}~\bibnamefont{Itano}},
  \bibinfo{author}{\bibfnamefont{S.}~\bibnamefont{Gilbert}}, \bibnamefont{and}
  \bibinfo{author}{\bibfnamefont{D.}~\bibnamefont{Wineland}},
  \bibinfo{journal}{Instrumentation and Measurement, IEEE Transactions on}
  \textbf{\bibinfo{volume}{40}}, \bibinfo{pages}{126} (\bibinfo{year}{1991}).

\bibitem[{\citenamefont{Kielpinski et~al.}(2002)\citenamefont{Kielpinski,
  Monroe, and Wineland}}]{Kielpinski2003}
\bibinfo{author}{\bibfnamefont{C.}~\bibnamefont{Kielpinski}},
  \bibinfo{author}{\bibfnamefont{C.}~\bibnamefont{Monroe}}, \bibnamefont{and}
  \bibinfo{author}{\bibfnamefont{D.}~\bibnamefont{Wineland}},
  \bibinfo{journal}{Nature} \textbf{\bibinfo{volume}{417}},
  \bibinfo{pages}{709} (\bibinfo{year}{2002}).

\bibitem[{\citenamefont{Monroe}(2002)}]{Monroe2002}
\bibinfo{author}{\bibfnamefont{C.}~\bibnamefont{Monroe}},
  \bibinfo{journal}{Nautre} \textbf{\bibinfo{volume}{416}},
  \bibinfo{pages}{238} (\bibinfo{year}{2002}).

\bibitem[{\citenamefont{King}(1999)}]{King1999}
\bibinfo{author}{\bibfnamefont{B.}~\bibnamefont{King}}, \bibinfo{journal}{PhD
  thesis} \textbf{\bibinfo{volume}{University of Colorado}}
  (\bibinfo{year}{1999}).

\bibitem[{\citenamefont{Leibfried et~al.}(2003)\citenamefont{Leibfried,
  DeMarco, Meyer, Lucas, Barrett, Britton, Itano, Jelenkovic, Langer, Rosenband
  et~al.}}]{DidiGate}
\bibinfo{author}{\bibfnamefont{D.}~\bibnamefont{Leibfried}},
  \bibinfo{author}{\bibfnamefont{B.}~\bibnamefont{DeMarco}},
  \bibinfo{author}{\bibfnamefont{V.}~\bibnamefont{Meyer}},
  \bibinfo{author}{\bibfnamefont{D.}~\bibnamefont{Lucas}},
  \bibinfo{author}{\bibfnamefont{M.}~\bibnamefont{Barrett}},
  \bibinfo{author}{\bibfnamefont{J.}~\bibnamefont{Britton}},
  \bibinfo{author}{\bibfnamefont{W.~M.} \bibnamefont{Itano}},
  \bibinfo{author}{\bibfnamefont{B.}~\bibnamefont{Jelenkovic}},
  \bibinfo{author}{\bibfnamefont{C.}~\bibnamefont{Langer}},
  \bibinfo{author}{\bibfnamefont{T.}~\bibnamefont{Rosenband}},
  \bibnamefont{et~al.}, \bibinfo{journal}{Nature}
  \textbf{\bibinfo{volume}{422}}, \bibinfo{pages}{412} (\bibinfo{year}{2003}).

\bibitem[{\citenamefont{Jost}(2010)}]{Jost2010}
\bibinfo{author}{\bibfnamefont{J.}~\bibnamefont{Jost}}, \bibinfo{journal}{PhD
  thesis} \textbf{\bibinfo{volume}{University of Colorado}}
  (\bibinfo{year}{2010}).

\bibitem[{\citenamefont{Biercuk et~al.}(2009)\citenamefont{Biercuk, Uys,
  VanDevender, Shiga, Itano, and Bollinger}}]{Biercuk2009I}
\bibinfo{author}{\bibfnamefont{M.}~\bibnamefont{Biercuk}},
  \bibinfo{author}{\bibfnamefont{H.}~\bibnamefont{Uys}},
  \bibinfo{author}{\bibfnamefont{A.}~\bibnamefont{VanDevender}},
  \bibinfo{author}{\bibfnamefont{N.}~\bibnamefont{Shiga}},
  \bibinfo{author}{\bibfnamefont{W.}~\bibnamefont{Itano}}, \bibnamefont{and}
  \bibinfo{author}{\bibfnamefont{J.~J.} \bibnamefont{Bollinger}},
  \bibinfo{journal}{Quantum Information and Computation}
  \textbf{\bibinfo{volume}{9}}, \bibinfo{pages}{920} (\bibinfo{year}{2009}).

\bibitem[{\citenamefont{Britton et~al.}(2012)\citenamefont{Britton, Sawyer,
  Keith, Wang, Freericks, Uys, Biercuk, and Bollinger}}]{Biercuk2012}
\bibinfo{author}{\bibfnamefont{J.}~\bibnamefont{Britton}},
  \bibinfo{author}{\bibfnamefont{B.}~\bibnamefont{Sawyer}},
  \bibinfo{author}{\bibfnamefont{A.}~\bibnamefont{Keith}},
  \bibinfo{author}{\bibfnamefont{C.}~\bibnamefont{Wang}},
  \bibinfo{author}{\bibfnamefont{J.}~\bibnamefont{Freericks}},
  \bibinfo{author}{\bibfnamefont{H.}~\bibnamefont{Uys}},
  \bibinfo{author}{\bibfnamefont{M.}~\bibnamefont{Biercuk}}, \bibnamefont{and}
  \bibinfo{author}{\bibfnamefont{J.}~\bibnamefont{Bollinger}},
  \bibinfo{journal}{Nature} \textbf{\bibinfo{volume}{484}},
  \bibinfo{pages}{489} (\bibinfo{year}{2012}).

\bibitem[{\citenamefont{Sawyer et~al.}(2012)\citenamefont{Sawyer, Britton,
  Kieth, Wang, Freericks, Uys, Biercuk, and Bollinger}}]{Sawyer2012}
\bibinfo{author}{\bibfnamefont{B.}~\bibnamefont{Sawyer}},
  \bibinfo{author}{\bibfnamefont{J.}~\bibnamefont{Britton}},
  \bibinfo{author}{\bibfnamefont{A.}~\bibnamefont{Kieth}},
  \bibinfo{author}{\bibfnamefont{C.}~\bibnamefont{Wang}},
  \bibinfo{author}{\bibfnamefont{J.}~\bibnamefont{Freericks}},
  \bibinfo{author}{\bibfnamefont{H.}~\bibnamefont{Uys}},
  \bibinfo{author}{\bibfnamefont{M.}~\bibnamefont{Biercuk}}, \bibnamefont{and}
  \bibinfo{author}{\bibfnamefont{J.}~\bibnamefont{Bollinger}},
  \bibinfo{journal}{Physical Review Letters} \textbf{\bibinfo{volume}{108}}
  (\bibinfo{year}{2012}).

\bibitem[{\citenamefont{Schnitzler et~al.}(2002)\citenamefont{Schnitzler,
  Frohlich, Boley, Clemen, Mlynek, Peters, and S.}}]{Schiller2002}
\bibinfo{author}{\bibfnamefont{H.}~\bibnamefont{Schnitzler}},
  \bibinfo{author}{\bibfnamefont{U.}~\bibnamefont{Frohlich}},
  \bibinfo{author}{\bibfnamefont{T.}~\bibnamefont{Boley}},
  \bibinfo{author}{\bibfnamefont{A.}~\bibnamefont{Clemen}},
  \bibinfo{author}{\bibfnamefont{J.}~\bibnamefont{Mlynek}},
  \bibinfo{author}{\bibfnamefont{A.}~\bibnamefont{Peters}}, \bibnamefont{and}
  \bibinfo{author}{\bibfnamefont{S.}~\bibnamefont{S.}}, \bibinfo{journal}{Appl.
  Opt.} \textbf{\bibinfo{volume}{41}}, \bibinfo{pages}{7000}
  (\bibinfo{year}{2002}).

\bibitem[{\citenamefont{Friedenauer et~al.}(2006)\citenamefont{Friedenauer,
  Markert, Schmitz, Petersen, Kahra, Herrmann, Udem, Hansch, and
  Schatz}}]{Hansch2006}
\bibinfo{author}{\bibfnamefont{A.}~\bibnamefont{Friedenauer}},
  \bibinfo{author}{\bibfnamefont{F.}~\bibnamefont{Markert}},
  \bibinfo{author}{\bibfnamefont{H.}~\bibnamefont{Schmitz}},
  \bibinfo{author}{\bibfnamefont{L.}~\bibnamefont{Petersen}},
  \bibinfo{author}{\bibfnamefont{S.}~\bibnamefont{Kahra}},
  \bibinfo{author}{\bibfnamefont{M.}~\bibnamefont{Herrmann}},
  \bibinfo{author}{\bibfnamefont{T.~H.} \bibnamefont{Udem}},
  \bibinfo{author}{\bibfnamefont{T.}~\bibnamefont{Hansch}}, \bibnamefont{and}
  \bibinfo{author}{\bibfnamefont{T.}~\bibnamefont{Schatz}},
  \bibinfo{journal}{Appl. Phys. B} \textbf{\bibinfo{volume}{84}},
  \bibinfo{pages}{371} (\bibinfo{year}{2006}).

\bibitem[{\citenamefont{Vasilyev et~al.}(2011)\citenamefont{Vasilyev, Nevsky,
  Ernsting, Hansen, Shen, and Schiller}}]{Schiller2011}
\bibinfo{author}{\bibfnamefont{S.}~\bibnamefont{Vasilyev}},
  \bibinfo{author}{\bibfnamefont{A.}~\bibnamefont{Nevsky}},
  \bibinfo{author}{\bibfnamefont{I.}~\bibnamefont{Ernsting}},
  \bibinfo{author}{\bibfnamefont{M.}~\bibnamefont{Hansen}},
  \bibinfo{author}{\bibfnamefont{J.}~\bibnamefont{Shen}}, \bibnamefont{and}
  \bibinfo{author}{\bibfnamefont{S.}~\bibnamefont{Schiller}},
  \bibinfo{journal}{Appl. Phys. B} \textbf{\bibinfo{volume}{103}},
  \bibinfo{pages}{27} (\bibinfo{year}{2011}).

\bibitem[{\citenamefont{Wilson et~al.}(2002)\citenamefont{Wilson, Ospelkaus,
  VanDevender, Mlynek, Brown, Leibfried, and Wineland}}]{Wilson2011}
\bibinfo{author}{\bibfnamefont{A.}~\bibnamefont{Wilson}},
  \bibinfo{author}{\bibfnamefont{C.}~\bibnamefont{Ospelkaus}},
  \bibinfo{author}{\bibfnamefont{A.}~\bibnamefont{VanDevender}},
  \bibinfo{author}{\bibfnamefont{J.}~\bibnamefont{Mlynek}},
  \bibinfo{author}{\bibfnamefont{K.}~\bibnamefont{Brown}},
  \bibinfo{author}{\bibfnamefont{D.}~\bibnamefont{Leibfried}},
  \bibnamefont{and} \bibinfo{author}{\bibfnamefont{J.}~\bibnamefont{Wineland}},
  \bibinfo{journal}{App. Phys. B} \textbf{\bibinfo{volume}{105}},
  \bibinfo{pages}{741 } (\bibinfo{year}{2002}).

\bibitem[{\citenamefont{Saleh and Tiech}(1991)}]{Saleh1991}
\bibinfo{author}{\bibfnamefont{B.}~\bibnamefont{Saleh}} \bibnamefont{and}
  \bibinfo{author}{\bibfnamefont{M.}~\bibnamefont{Tiech}},
  \emph{\bibinfo{title}{Fundamentals of Photonics}} (\bibinfo{publisher}{John
  Wiley \& Sons, Inc}, \bibinfo{address}{New York}, \bibinfo{year}{1991}).

\bibitem[{\citenamefont{Siegman}(1986)}]{Siegman1986}
\bibinfo{author}{\bibfnamefont{A.}~\bibnamefont{Siegman}},
  \emph{\bibinfo{title}{Lasers}} (\bibinfo{publisher}{Oxford University Press},
  \bibinfo{address}{Oxford}, \bibinfo{year}{1986}).

\bibitem[{\citenamefont{Ricci et~al.}(1995)\citenamefont{Ricci, Weidemiiller,
  Esslinger, Hemmerich, Zimmermann, Vuletic, Kijnig, and Hansch}}]{Ricci1995}
\bibinfo{author}{\bibfnamefont{L.}~\bibnamefont{Ricci}},
  \bibinfo{author}{\bibfnamefont{M.}~\bibnamefont{Weidemiiller}},
  \bibinfo{author}{\bibfnamefont{T.}~\bibnamefont{Esslinger}},
  \bibinfo{author}{\bibfnamefont{A.}~\bibnamefont{Hemmerich}},
  \bibinfo{author}{\bibfnamefont{C.}~\bibnamefont{Zimmermann}},
  \bibinfo{author}{\bibfnamefont{V.}~\bibnamefont{Vuletic}},
  \bibinfo{author}{\bibfnamefont{W.}~\bibnamefont{Kijnig}}, \bibnamefont{and}
  \bibinfo{author}{\bibfnamefont{T.}~\bibnamefont{Hansch}},
  \bibinfo{journal}{Optics Communications} \textbf{\bibinfo{volume}{117}},
  \bibinfo{pages}{541} (\bibinfo{year}{1995}).

\bibitem[{\citenamefont{Fan}(1951)}]{Fan1951}
\bibinfo{author}{\bibfnamefont{H.}~\bibnamefont{Fan}}, \bibinfo{journal}{Phys.
  Rev.} \textbf{\bibinfo{volume}{82}}, \bibinfo{pages}{900}
  (\bibinfo{year}{1951}).

\bibitem[{\citenamefont{Varshni}(1967)}]{Varshni1967}
\bibinfo{author}{\bibfnamefont{Y.}~\bibnamefont{Varshni}},
  \bibinfo{journal}{Physica} \textbf{\bibinfo{volume}{34}},
  \bibinfo{pages}{149} (\bibinfo{year}{1967}).

\bibitem[{\citenamefont{O'Donnell and Chen}(1991)}]{Chen1991}
\bibinfo{author}{\bibfnamefont{K.}~\bibnamefont{O'Donnell}} \bibnamefont{and}
  \bibinfo{author}{\bibfnamefont{X.}~\bibnamefont{Chen}},
  \bibinfo{journal}{Appl. Phys. Lett.} \textbf{\bibinfo{volume}{58}},
  \bibinfo{pages}{2924} (\bibinfo{year}{1991}).

\bibitem[{\citenamefont{Wieman and Hollberg}(1991)}]{Wieman1991}
\bibinfo{author}{\bibfnamefont{C.}~\bibnamefont{Wieman}} \bibnamefont{and}
  \bibinfo{author}{\bibfnamefont{L.}~\bibnamefont{Hollberg}},
  \bibinfo{journal}{Rev. Sci. Instrum.} \textbf{\bibinfo{volume}{62}},
  \bibinfo{pages}{1} (\bibinfo{year}{1991}).

\bibitem[{\citenamefont{MacAdam et~al.}(1992)\citenamefont{MacAdam, Steinbach,
  and Wieman}}]{MacAdam1992}
\bibinfo{author}{\bibfnamefont{K.}~\bibnamefont{MacAdam}},
  \bibinfo{author}{\bibfnamefont{A.}~\bibnamefont{Steinbach}},
  \bibnamefont{and} \bibinfo{author}{\bibfnamefont{C.}~\bibnamefont{Wieman}},
  \bibinfo{journal}{American journal of Physics} \textbf{\bibinfo{volume}{60}},
  \bibinfo{pages}{1098} (\bibinfo{year}{1992}).

\bibitem[{\citenamefont{Hecht}(1974)}]{Hecht1974}
\bibinfo{author}{\bibfnamefont{U.}~\bibnamefont{Hecht}},
  \emph{\bibinfo{title}{Optics}} (\bibinfo{publisher}{Addison Wesley},
  \bibinfo{address}{Reading, Massechusetts}, \bibinfo{year}{1974}).

\bibitem[{Tie()}]{Tiecke_thesis}
\bibinfo{note}{T. Tiecke, \emph{PhD Thesis, University of Amsterdam}, (2009)}.

\bibitem[{not()}]{note1}
\bibinfo{note}{S. Gensemer acknowledges the assistance of T. Tiecke in the
  design of the hermetically sealed diode laser mount.}

\bibitem[{\citenamefont{Kato et~al.}()\citenamefont{Kato, Baba, Kasahara,
  Ishikawa, Nisono, Kimura, O'Reilly, Kuwano, Shimamoto, Shinano
  et~al.}}]{IodineAtlas}
\bibinfo{author}{\bibfnamefont{H.}~\bibnamefont{Kato}},
  \bibinfo{author}{\bibfnamefont{M.}~\bibnamefont{Baba}},
  \bibinfo{author}{\bibfnamefont{S.}~\bibnamefont{Kasahara}},
  \bibinfo{author}{\bibfnamefont{K.}~\bibnamefont{Ishikawa}},
  \bibinfo{author}{\bibfnamefont{M.}~\bibnamefont{Nisono}},
  \bibinfo{author}{\bibfnamefont{Y.}~\bibnamefont{Kimura}},
  \bibinfo{author}{\bibfnamefont{J.}~\bibnamefont{O'Reilly}},
  \bibinfo{author}{\bibfnamefont{H.}~\bibnamefont{Kuwano}},
  \bibinfo{author}{\bibfnamefont{T.}~\bibnamefont{Shimamoto}},
  \bibinfo{author}{\bibfnamefont{T.}~\bibnamefont{Shinano}},
  \bibnamefont{et~al.}, \bibinfo{journal}{Faculty of Science, Kobe University,
  Nada-ku, Kobe, Japan}  (????).

\bibitem[{\citenamefont{Ratnapala et~al.}(2004)\citenamefont{Ratnapala, Vale,
  White, Harvey, Heckenberg, and Rubinsztein-Dunlop}}]{Ratnapala2004}
\bibinfo{author}{\bibfnamefont{A.}~\bibnamefont{Ratnapala}},
  \bibinfo{author}{\bibfnamefont{C.}~\bibnamefont{Vale}},
  \bibinfo{author}{\bibfnamefont{A.}~\bibnamefont{White}},
  \bibinfo{author}{\bibfnamefont{M.}~\bibnamefont{Harvey}},
  \bibinfo{author}{\bibfnamefont{N.}~\bibnamefont{Heckenberg}},
  \bibnamefont{and}
  \bibinfo{author}{\bibfnamefont{H.}~\bibnamefont{Rubinsztein-Dunlop}},
  \bibinfo{journal}{Optics Letters} \textbf{\bibinfo{volume}{29}},
  \bibinfo{pages}{2704} (\bibinfo{year}{2004}).

\end{thebibliography}

\end{document}